# A Novel approach as Multi-place Watermarking for Security in Database

Brijesh B. Mehta, Udai Pratap Rao
Dept. of Computer Engineering, S. V. National Institute of Technology, Surat, Gujarat, INDIA-395007

**Abstract-** *Digital multimedia watermarking technology had suggested in the last decade to embed copyright information in digital objects such as images, audio and video. However, the increasing use of relational database systems in many real-life applications created an ever-increasing need for watermarking database systems. As a result, watermarking relational database systems is now merging as a research area that deals with the legal issue of copyright protection of database systems. The main goal of database watermarking is to generate robust and impersistant watermark for database. In this paper we propose a method, based on image as watermark and this watermark is embedded over the database at two different attribute of tuple, one in the numeric attribute of tuple and another in the date attribute's time (seconds) field. Our approach can be applied for numerical and categorical database.*

## 1 Introduction

Watermarking is firstly, introduced for image processing and then it extended to security of text and multimedia data. Now days, it also used for database and software. There is so much work done so far by many researches in the watermarking multimedia data [1] [2] [3]. Most of this method were initially developed for images [4] and later extended to video [5] and audio data [6][7]. Software watermarking techniques [8][9][10][11], also been introduced but it did not get much success because they are easily detectable in code. Due to differences between multimedia and database we cannot directly use any of the technique as it is for database, which developed for multimedia data. These differences include [12][13]:

- A multimedia object consists of a large number of bits, with considerable redundancy. Therefore, the watermark has more space to hide where as a database relation consists of tuples, each of which represents a separate object. So the watermark needs to be spread over these separate objects.
- The relative spatial/temporal positioning of various pieces of a multimedia object typically does not change. Whereas, tuples may changes with updates in database.
- Portions of a multimedia object cannot be dropped or replaced arbitrarily without causing perceptual changes in the object. Whereas, tuples may simply be dropped by delete operation in database.

In this paper, we have proposed a new approach for robust database watermarking.[15] When we are talking about watermarking as copyright protection robustness is a very important issue as to prove ownership of data, user have to detect their watermark in data without any damage or say defect this defines robustness of watermark. Means the rate of correctly detection of watermark is also called robustness of the watermark. There are two main stages when we apply watermarking to data, (1) Watermark Embedding, and (2) Watermark Detection. In Watermark Embedding or say watermark insertion, we are applying or inserting a watermark in to the object or data, which, we want to protect. In watermark detection or watermark extraction, we try to extract the watermark from data or object or just check for the presence of watermark in data in some cases.

Rest of the paper is organized as follows: in Sec. 2, we discuss related work in this area. In Sec. 3, we discuss overview of our novel approach. In Sec. 4, we discuss algorithm for our approach. In Sec. 5, we evaluated the performance of our algorithm with reference to different attacks. Then conclusion and future work in Sec. 6 and references in Sec. 7.

## 2 Related Work

Watermarking relational databases is a relatively new research area that deals with the legal issue of copyright protection of relational databases. Therefore, literature in this area has been very limited, and focused on embedding short strings of binary bits in numerical databases [18]. In year 2000, S. Khanna et al. proposed the novel idea of controlling the security of database with digital watermark [14], which arouses the researchers' interest in watermarking database. So mainly there are two paths come out for database watermarking.

Firstly Agrawal et al.[12] presented a scheme and implemented it. This algorithm assumes that numeric attribute can tolerate modifications of some least significant bits. So, Tuples selected first for watermark embedding. Then certain bits of some attributes of the selected tuples modified to embed watermark bits.

Second scheme of sion et al.[15] in which, all tuples securely divided into non-intersecting subsets. A single watermark bit embedded into tuples of a subset by modifying the distribution of tuples values. The same watermark bit embedded repeatedly across several subsets and the majority voting technique employed to detect the watermark.

Some other authors had also tried to improve above two approaches with their own ideas to make them more secure and robust. From these we have studied some of the papers which are:

- Watermark based copyright protection of outsourced database by ZHU Qin et al.[16] Which, explains a database watermarking method based on first approach we discussed above with chaotic random number generator.
- A Speech based algorithm for watermarking relational databases by haiqing wang et al. [17] which, explains a database watermarking methods same as above but here they have used voice as a watermark.
- Watermarking Relational Database Systems by Ashraf Odeh et al.[18] which, explains a watermarking method based on first approach but here image is used as watermark and it is embedded in Date attribute's time field. We have also used this algorithm in our approach as one of the algorithm to embed watermark.
- Robust and Blind Watermarking of Relational Database Systems by Ali Al-Haj et al.[19] which, explains image based watermark embedding method in the non-numeric field of database as spaces and double space for encoding the watermark bit.
- One of the latest paper we go through is about a new relational watermarking scheme resilient to additive attacks by Nagarjuna Settipalli & R. Manjula.[20] As title suggest they just taken care of additive attacks applied on database watermarking.

## 3    Proposed Approach

Our proposed approach is based on the modification of two algorithms given by Agrawal et al. [13] And Ashraf Odeh et al. [18]. In earlier approach[13][18] author used a single attribute of a tuple to embed a watermark but we are embedding the same watermark in two attributes using our proposed algorithms. Therefore, it will be difficult for attacker to remove both watermarks from the database, based on extracted bits from both algorithms we can generate original watermark very easily, and we can prove ownership of database. In our method binary image is used as watermark. The whole procedure of embedding and extraction of watermark is performing in two phases, in first phase we insert watermark in the numeric field of the database and in second phase we insert a watermark in the seconds field of database same way at the time of extraction we follow the reverse order of above phases.

## 4    Algorithms

Main purpose of writing this algorithm is to give more robust database watermarking technique. In earlier techniques they insert watermark at one place only so we found that, it can be removed by some of the database update operation where is in our approach its little more difficult.

Notations used in this algorithm are:
- $n$ - Number of tuples in the relation
- $v$ - Number of attributes in the relation available for marking
- $k_1$(5-bits)- Key used to determine the place where watermark can be inserted
- $k_2$(4-bits)- Key used to generate watermark for second phase

### 4.1    Watermark insertion / embedding

Suppose that the scheme of a database relation is R(P, $A_0$, ... , $A_{v-1}$), where P is primary key attribute, If there is no primary key, auto-increasing attribute will be added to act as primary key. Assume that all $v$ attribute are numeric and are candidates for marking.

A very important assumption regarding database watermarking is that small changes in LSB of a numeric attribute are tolerable within certain precision range. To get a copyright protection data owner should pay this price. In fact, it is noteworthy that the publisher of books of mathematical tables has been introducing small errors in their tables for centuries to identify pirated copies [3].

Here we are going to use two keys for the watermark insertion $k_1$ and $k_2$. Both $k_1$ and $k_2$ are known by database owner only. $k_1$ is used to select the tuple in which we need to insert a watermark where as $k_2$ is a 4 bit key which is used to convert single bit watermark into 5 bit watermark to easily embedding it into time field of date attribute. The parameters $v$, $n$, $k_1$, $k_2$ are known to the owner only. We are using binary image as watermark.

In first phase we embed the watermark using the modified algorithm of Agrawal et al.[13] we have removed a keyed hash function from the algorithm provided by Agrawal et al. and made it simple to insert and detect though we compromise with the security but as we are inserting watermark at two places, we can take this chances.

1. For each tuple r ∈ R do
2.   If (F(r.P) mod k1 equals 0) then

3. For each attribute of tuple
4.   If(attribute ϵ *v*) then
5.    Find LSB of that attribute and replace it with watermark bit
6.   End if
7.  End loop
8. End if
9. End loop

Therefore, by this way we have inserted watermark at one place now it's time to insert the same watermark at other place and that other place is the time field of date attribute. Hiding the binary information in the seconds field (SS) should have the least effect on the usability of the database. A major advantage of using the time attribute is the large bit-capacity available for hiding the watermark, and thus large watermarks can be easily hide, if required. Therefore, in second phase we use algorithm proposed by Ashraf et al. [18] with slight modification as we are taking MM (Minuit) field to decide that is it possible to insert a watermark or not. Whereas in algorithm given by Ashraf et al., they are using SS field and directly inserting 5 bits of binary image into the SS field whereas we are inserting only 1 bit of image which is concatenated with 4 bit of key. As we are embedding single bit in attribute but this algorithm uses 5-bit watermark so we first do concatenation operation between key $k_2$ and watermark bit to get 5 bit new watermark and then uses the algorithm to insert a watermark.

1. Concatenation of the value of watermark bit and $k_2$
2. Find the decimal equivalent of the string
3. Embed the decimal number in tuples selected by the pre-defined key $k_1$ as follows:
   3.1. For each selected tuple do
   3.2.   For each selected Time attribute do
   3.3.    If the 'MM' field of the 'Time' mode $k_1 = 0$
   3.4.      Embed the decimal number in SS field
   3.5.    Else Next attribute
   3.6.    End if
   3.7.   End loop
   3.8. End loop

Therefore, by this way we have completed multi-place watermark insertion in the database but now the difficult part is to extract the watermark from two places and then comparing them to check for the original watermark

### 4.2 Watermark extraction

Watermark Detection is the procedure to check whether watermark is present in data where as in watermark extraction watermark is been carried out and regenerated.–Algorithm designed by Agrawal et al.[13] was for detection of the watermark not to extract it because it calculates total count and match count, so we need to slightly modify it to extract the watermark. So, as we mention earlier in insertion that we have removed the keyed hash function from the algorithm and we are actually taking values from the data not only checking that whether watermark is there or not.

1. For each tuple s ϵ S do
2. If((s.P) mod k1 equals 0) then
3.   For each atribute of tuple
4.    If(attribute ϵ *v*) then
5.     Find LSB of that attribute and extract it as our watermark
6.    End if
7.   End loop
8. End if
9. End loop

Now, in above algorithm if we find any bit that has changed after insertion, then we are putting it as zero and all detected bits as it is. So, if we find 1 in the watermark then we can say that they are the correct values but for zero we can't say anything yet but after second phase we have more clear idea about the original watermark.

In second phase, we are using Ashraf Odeh et al.'s algorithm with slight modification that instead of generating binary image from that binary equivalent of the extracted watermark. We are taking only LSB of the binary data.

1. Extract the decimal number in tuple selected by the pre-defined key $k_1$ as follows:
   1.1. For each selected tuple do
   1.2.   For each selected 'Time' attribute do
   1.3.    If the 'MM' field of the 'time' mode $k_1 = 0$
   1.4.      Extract the decimal number from SS field
   1.5.    Else Next attribute
   1.6.    End if
   1.7.   End loop
   1.8. End loop
2. Find the binary equivalent of the extracted decimal number
3. Extract last bit (LSB) from is which indicate our watermark.

Now, we have two watermarks, which may be changed or not changed. Now, we will compare these two watermarks with original watermark. Therefore, if any one of them will match we will consider it true and by this way we can get the correctly extracted watermark in percentage by the following algorithm.

```
Matchcount=0, totalcount=0
For each bit of watermark
  If( WM = WM1 or WM=WM2) then
    matchcount=matchcount+1
  End if
  Totalcount=totalcount+1
End for
```

## 5  Performance Evaluation

Proposed algorithm has been tested and evaluated on an experimental data set consists of approximately 5000 tuples. We are taking care of robustness in our performance evaluation. We have completed all our programming in C language. We have considered few of the major attacks applied on Database Watermarking like:

### 5.1  Subset addition attack

In this type of attack, the attacker adds a set of tuples to the original database.[18] This kind of attack does not have any effect on our algorithm coz new tuples added have no watermark embedded in them so at the time of extraction they are simply ignored so we get 100% extraction after adding 100% new tuples.

### 5.2  Subset deletion attack

In this type of attack, the attacker may delete a subset of the tuples of the watermarked database hoping that the watermark will be removed.[18] Graph in Figure 1 shows that watermark can completely only be removed by all the tuples because if 5% tuples are there then also we can extract from the database this shows that our algorithm is very robust against this kind of attack.

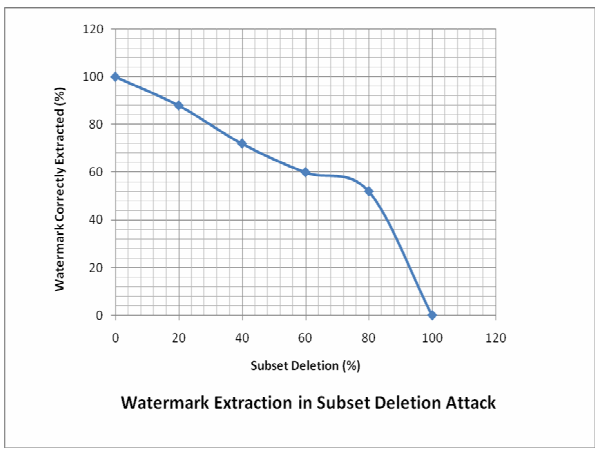

Fig. 1. Robustness Results due to the Subset Deletion Attack

### 5.3  Subset alteration attack

In this type of attack, the attacker alters the tuples of the database through operations such as linear transformation. [18] By doing this attacker thinks that he/she will be get success in removing the watermark from database but graph in figure 2 shows that even altering all the tuples watermark can still be extracted.

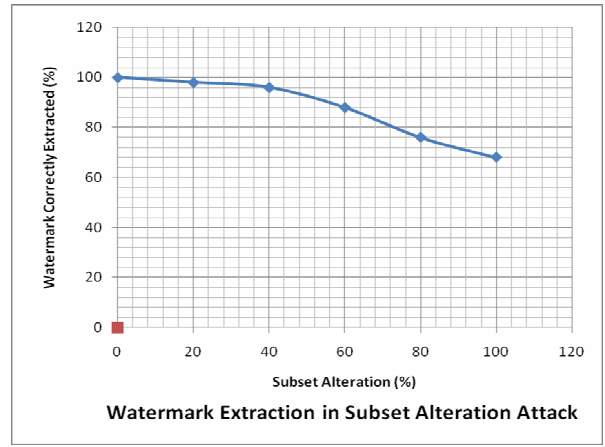

Fig. 2. Robustness Results due to the Subset Alteration Attack

### 5.4  Subset selection attack

In this type of attack, the attacker randomly selects a subset of the original database that might still provide value for its intended purpose. [18] By doing this attacker thinks that the small he selects has no watermark embedded in it but as our algorithm embeds watermark at two places even he selects 10% part of dataset it contains a watermark in it and it can be proved by the graph in the figure 3.

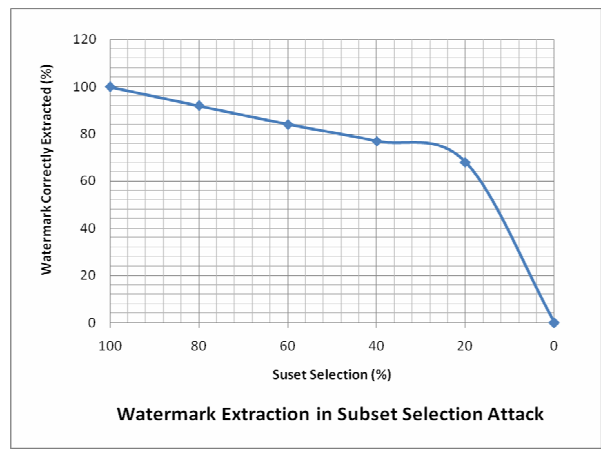

Fig. 3. Robustness Results due to the Subset selection Attack

## 6  Conclusions and Future Work

We have proposed a novel approach for robust database watermarking which is very useful in copyright protection of database. As we are inserting same watermark at different places so, there is a less chances of it to get attacked and if so, it is comparatively easy to extract the original watermark because of the watermark is embedded at two places. We can have one correctly extracted watermark from that two places. Though it may be more costlier in a context of data correctness as it changes many attributes of a tuple in database but if we are thinking about robust copyright protection then we need to pay this price.

As in our proposed approach we are not only detecting the watermark but also extracting it. Therefore, instead of binary image we can insert some kind of biometrics data like, speech. For eg. Speech has a unique characteristic means almost everyone's voice is different than the other person.